\documentclass[prl,twocolumn,lengthcheck,superscriptaddress,letterpaper,nofootinbib,nopreprintnumbers,showpacs]{revtex4}

\usepackage{epsfig}
\usepackage{graphicx}
\usepackage{color}
\usepackage{ulem}
\usepackage{endnotes}
\usepackage{amssymb}
\usepackage{amsmath,bm}
\usepackage{times}
\usepackage{subfigure}
\usepackage{verbatim}

\newcommand{\be}{\begin{equation}}
\newcommand{\ee}{\end{equation}}
\newcommand{\bdm}{\begin{displaymath}}
\newcommand{\edm}{\end{displaymath}}
\newcommand{\bea}{\begin{eqnarray}}
\newcommand{\eea}{\end{eqnarray}}

\newcommand{\mtrx}[1]{\bm{#1}}

\def\lsim{\lower.5ex\hbox{$\; \buildrel < \over \sim \;$}}

\newcommand{\incgraph}[3]{\includegraphics[angle=#1, width=#2\textwidth]{#3}}

\begin{document}

%\title{European Pulsar Timing Array constraints on anisotropy in the nanohertz stochastic gravitational-wave background}
\title{Limits on anisotropy in the nanohertz stochastic gravitational-wave background}


\author{S. R. Taylor}
\email{Stephen.R.Taylor@jpl.nasa.gov}
\affiliation{Jet Propulsion Laboratory, California Institute of Technology, Pasadena, California 91109, USA}
\affiliation{Institute of Astronomy, University of Cambridge,
  Madingley Road, Cambridge CB3 0HA, UK}

\author{C. M. F. Mingarelli}
\affiliation{TAPIR (Theoretical Astrophysics), California Institute of Technology MC 350-17, Pasadena, California 91125, USA}
\affiliation{Max-Planck-Institut f{\"u}r Radioastronomie, Auf dem H{\"u}gel 69, D-53121 Bonn, Germany}
\affiliation{School of Physics and Astronomy, University of
  Birmingham, Edgbaston, Birmingham B15 2TT, United Kingdom}

\author{J. R. Gair}
\affiliation{Institute of Astronomy, University of Cambridge,
  Madingley Road, Cambridge CB3 0HA, UK}
\author{A. Sesana}
\affiliation{School of Physics and Astronomy, University of
  Birmingham, Edgbaston, Birmingham B15 2TT, United Kingdom}
\affiliation{Max-Planck-Institut f{\"u}r Gravitationsphysik, Albert
  Einstein Institut, Am M{\"u}hlenberg 1, 14476 Golm, Germany}
\author{G. Theureau}
\affiliation{Laboratoire de Physique et Chimie de l'Environnement et
  de l'Espace LPC2E CNRS-Universit{\'e} d'Orl{\'e}ans, F-45071
  Orl{\'e}ans, France}
\affiliation{Station de radioastronomie de Nan{\c c}ay, Observatoire
  de Paris, CNRS/INSU F-18330 Nan{\c c}ay, France}
\affiliation{Laboratoire Univers et Th{\'e}ories LUTh, Observatoire de Paris, CNRS/INSU, Universit\'e Paris Diderot, 5 place Jules Janssen, 92190 Meudon, France}

\author{S. Babak}
\affiliation{Max-Planck-Institut f{\"u}r Gravitationsphysik, Albert
  Einstein Institut, Am M{\"u}hlenberg 1, 14476 Golm, Germany}
\author{C. G. Bassa}
\affiliation{ASTRON, the Netherlands Institute for Radio Astronomy,
  Postbus 2, 7990 AA, Dwingeloo, The Netherlands}
\affiliation{Jodrell Bank Centre for Astrophysics, University of
  Manchester, Manchester, M13 9PL, United Kingdom}
\author{P. Brem}
\affiliation{Max-Planck-Institut f{\"u}r Gravitationsphysik, Albert
  Einstein Institut, Am M{\"u}hlenberg 1, 14476 Golm, Germany}
\author{M. Burgay}
\affiliation{INAF - Osservatorio Astronomico di Cagliari, via della
  Scienza 5, I-09047 Selargius (CA), Italy}
\author{R. N. Caballero}
\affiliation{Max-Planck-Institut f{\"u}r Radioastronomie, Auf dem H{\"u}gel 69, D-53121 Bonn, Germany}
\author{D. J. Champion}
\affiliation{Max-Planck-Institut f{\"u}r Radioastronomie, Auf dem H{\"u}gel 69, D-53121 Bonn, Germany}
\author{I. Cognard}
\affiliation{Laboratoire de Physique et Chimie de l'Environnement et
  de l'Espace LPC2E CNRS-Universit{\'e} d'Orl{\'e}ans, F-45071
  Orl{\'e}ans, France}
\affiliation{Station de radioastronomie de Nan{\c c}ay, Observatoire
  de Paris, CNRS/INSU F-18330 Nan{\c c}ay, France}
\author{G. Desvignes}
\affiliation{Max-Planck-Institut f{\"u}r Radioastronomie, Auf dem
  H{\"u}gel 69, D-53121 Bonn, Germany}
\author{L. Guillemot}
\affiliation{Laboratoire de Physique et Chimie de l'Environnement et
  de l'Espace LPC2E CNRS-Universit{\'e} d'Orl{\'e}ans, F-45071
  Orl{\'e}ans, France}
\affiliation{Station de radioastronomie de Nan{\c c}ay, Observatoire
  de Paris, CNRS/INSU F-18330 Nan{\c c}ay, France}
\author{J. W. T. Hessels}
\affiliation{ASTRON, the Netherlands Institute for Radio Astronomy,
  Postbus 2, 7990 AA, Dwingeloo, The Netherlands}
\affiliation{Anton Pannekoek Institute for Astronomy, University of
  Amsterdam, Science Park 904, 1098 XH Amsterdam, The Netherlands}
\author{G. H. Janssen}
\affiliation{ASTRON, the Netherlands Institute for Radio Astronomy,
  Postbus 2, 7990 AA, Dwingeloo, The Netherlands}
\affiliation{Jodrell Bank Centre for Astrophysics, University of
  Manchester, Manchester, M13 9PL, United Kingdom}
\author{R. Karuppusamy}
\affiliation{Max-Planck-Institut f{\"u}r Radioastronomie, Auf dem
  H{\"u}gel 69, D-53121 Bonn, Germany}
\author{M. Kramer}
\affiliation{Max-Planck-Institut f{\"u}r Radioastronomie, Auf dem
  H{\"u}gel 69, D-53121 Bonn, Germany}
\affiliation{Jodrell Bank Centre for Astrophysics, University of
  Manchester, Manchester, M13 9PL, United Kingdom}
\author{A. Lassus}
\affiliation{Max-Planck-Institut f{\"u}r Radioastronomie, Auf dem
  H{\"u}gel 69, D-53121 Bonn, Germany}
\affiliation{Laboratoire de Physique et Chimie de l'Environnement et
  de l'Espace LPC2E CNRS-Universit{\'e} d'Orl{\'e}ans, F-45071
  Orl{\'e}ans, France}
\author{P. Lazarus}
\affiliation{Max-Planck-Institut f{\"u}r Radioastronomie, Auf dem
  H{\"u}gel 69, D-53121 Bonn, Germany}
\author{L. Lentati}
\affiliation{Astrophysics Group, Cavendish Laboratory, JJ Thomson Avenue, Cambridge, CB3 0HE, UK}
\author{K. Liu}
\affiliation{Max-Planck-Institut f{\"u}r Radioastronomie, Auf dem
  H{\"u}gel 69, D-53121 Bonn, Germany}
\author{S. Os{\l}owski}
\affiliation{Fakult\"at f\"ur Physik, Universit\"at Bielefeld,
  Postfach 100131, 33501 Bielefeld, Germany}
  \affiliation{Max-Planck-Institut f{\"u}r Radioastronomie, Auf dem
  H{\"u}gel 69, D-53121 Bonn, Germany}
\author{D. Perrodin}
\affiliation{INAF - Osservatorio Astronomico di Cagliari, via della
  Scienza 5, I-09047 Selargius (CA), Italy}
\author{A. Petiteau}
\affiliation{Universit\'e Paris-Diderot-Paris7 APC - UFR de Physique,
  Batiment Condorcet, 10 rue Alice Domont et L\'eonie Duquet 75205
  PARIS CEDEX 13, France}
\author{A. Possenti}
\affiliation{INAF - Osservatorio Astronomico di Cagliari, via della
  Scienza 5, I-09047 Selargius (CA), Italy}
\author{M. B. Purver}
\affiliation{Jodrell Bank Centre for Astrophysics, University of
  Manchester, Manchester, M13 9PL, United Kingdom}
\author{P. A. Rosado}
\affiliation{Centre for Astrophysics \& Supercomputing, Swinburne
  University of Technology, PO Box 218, Hawthorn VIC 3122, Australia}
\affiliation{Max Planck Institute for Gravitational Physics, Albert
  Einstein Institute, Callinstra\ss e 38, 30167, Hanover, Germany}
\author{S.~A.~Sanidas}
\affiliation{Jodrell Bank Centre for Astrophysics, University of Manchester, Manchester, M13 9PL, United Kingdom}
\affiliation{Anton Pannekoek Institute for Astronomy, University of
  Amsterdam, Science Park 904, 1098 XH Amsterdam, The Netherlands}
\author{R. Smits}
\affiliation{ASTRON, the Netherlands Institute for Radio Astronomy,
  Postbus 2, 7990 AA, Dwingeloo, The Netherlands}
\author{B. Stappers}
\affiliation{Jodrell Bank Centre for Astrophysics, University of
  Manchester, Manchester, M13 9PL, United Kingdom}
\author{C. Tiburzi}
\affiliation{INAF - Osservatorio Astronomico di Cagliari, via della
  Scienza 5, I-09047 Selargius (CA), Italy}
\affiliation{Dipartimento di Fisica - Universit\'a di Cagliari,
  Cittadella Universitaria, I-09042 Monserrato (CA), Italy}
\author{R. van Haasteren}
\affiliation{Jet Propulsion Laboratory, California Institute of
  Technology, Pasadena, California 91109, USA}
\author{A. Vecchio}
\affiliation{School of Physics and Astronomy, University of
  Birmingham, Edgbaston, Birmingham B15 2TT, United Kingdom}
\author{J. P. W. Verbiest}
\affiliation{Fakult\"at f\"ur Physik, Universit\"at Bielefeld,
  Postfach 100131, 33501 Bielefeld, Germany}
  \affiliation{Max-Planck-Institut f{\"u}r Radioastronomie, Auf dem
  H{\"u}gel 69, D-53121 Bonn, Germany}

\date\today

\begin{abstract}
The paucity of observed supermassive black hole binaries (SMBHBs) may
imply that the gravitational wave background (GWB) from this population is anisotropic, rendering existing analyses
sub-optimal. We present the first constraints on the
angular distribution of a nanohertz stochastic GWB from
circular, inspiral-driven SMBHBs using the $2015$ European Pulsar
Timing Array data~\citep{paperI}. 
Our analysis of the GWB in the $\sim 2 - 90$~nHz band shows consistency with isotropy, with the strain amplitude in $l>0$ spherical harmonic multipoles $\lesssim
40\%$ of the monopole value. We expect that these more general techniques will become standard
tools to probe the angular distribution of source populations. 
\end{abstract}


\maketitle

{\it Introduction.}-- Pulsar Timing Arrays (PTAs) are currently being
used to search for, and to eventually characterize, the nanohertz
stochastic gravitational-wave background (SGWB) by looking for
  correlated deviations in the pulse times of arrival (TOAs) of
  multiple radio millisecond pulsars distributed across the sky. The
  SGWB in the nanohertz regime is thought to be generated by the incoherent superposition of a large number of weak and unresolved GW sources, including supermassive black hole binaries (SMBHBs) \citep{RajagopalRomani:1995, wyithe-loeb-2003, JaffeBacker:2003,
  sesana-vecchio-volonteri-2009,
  sesana-vecchio-colacino-2008,wenjenet2011, sesana-review-2013-1},
decaying cosmic-string networks
\citep{vilenkin-1981a,vilenkin-1981b,olmez-2010,sbs2012} or primordial
GWs \citep{grishchuk-1976,grishchuk-2005}. Previous analyses have
assumed background isotropy, which emerges as a special case from the more general
anisotropy framework presented here.
Although GWs have not yet been directly detected,
limits on the angular power distribution of a nanohertz SGWB may constrain
  the distribution of low redshift structure \citep{rs2014}, the location of
  several particularly
  bright nearby sources dominating the signal strain budget \citep{rsg2015,ravi-2012}, and
  open a new avenue to explore the population characteristics of
  SMBHBs. Moreover, if a significant
  fraction of SMBHBs stall rather than merge, or are rapidly driven to
  merger via strong couplings to the galactic nuclear environment,
  then we may expect a depleted nanohertz GW signal dominated by only a few
  bright sources \citep{ks2011}. As such, the tools implemented here
  may provide new and novel insights into the final-parsec problem
  (see e.g. Ref.\ \citep{finalparsecproblem2003}). This research is a result of the common effort to directly detect gravitational waves using pulsar timing, known as the European Pulsar Timing Array
\citep[EPTA,][]{2013CQGra..30v4009K}.

Limits on the SGWB are usually reported in terms of the characteristic-strain spectrum $h_c(f)$ of a background which is composed of purely
GW-driven, circular, inspiraling SMBHBs which obeys a simple
power-law: 
\begin{equation}
\label{eq:hc}
h_c(f) = A_h(f/\text{yr}^{-1})^{-2/3} \, ,
\end{equation}
where $A_h$ is the
strain amplitude reported at a reference frequency $f=\mathrm{yr}^{-1}$ \cite{phinney:2001}.
The correlations induced by a SGWB in pulsar TOAs can be understood by considering a perturbation to the space-time metric 
along the Earth-pulsar line of sight causing a change in the perceived rotational-frequency of the pulsar \citep{sazhin-1978,detweiler-1979,estabrook-1975,burke-1975}. The fractional frequency shift $\delta \nu (t)/\nu_0$ of a signal from a pulsar at rest frequency $\nu_0$ is 
the difference in the metric perturbation at the Solar System barycenter (SSB), 
and at the pulsar.
This frequency shift is integrated over time to give the induced timing residuals, $r(t)\equiv\int^t \delta\nu(t')/\nu_0dt'$, which
are cross-correlated between pulsars in an effort to boost the
detection probability of GW signals at the
Earth. The expectation value of the cross-correlated
timing residuals between pulsars $a$ and $b$ is proportional to the overlap reduction function
(ORF, $\Gamma_{ab}$)-- a dimensionless
function which quantifies the response of a pair of pulsars to the
stochastic GW background \cite{Finn:2009,MingarelliSidery:2014}. In this Letter, we use
analytically computed anisotropic ORFs
\citep{MingarelliEtAl:2013,GairEtAl:2014} and recently developed
Bayesian techniques \citep{TaylorGair:2013} to
constrain the angular power distribution of the SGWB.

{\it Fitting a pulsar timing model.}--
The average pulse profiles of
millisecond pulsars are
remarkably stable and reproducible.
This stability permits high-precision timing, which is
crucial to GW searches: the minimum detectable GW strain is $h_c \propto 10^{-15}
(\sigma_{\rm rms}/100~\mathrm{ns})(T/10~\mathrm{yr})^{-1}$, where
$\sigma_{\rm rms}$
is the root-mean-square of the pulsar timing residuals, and $T$ is the total observation
timespan \citep{sesana-vecchio-colacino-2008}. Therefore high timing precision and long-term observations are
required to distinguish the GW signal from
noise, as well as
boost the signal-to-noise ratio (SNR) in a search.

Pulsar observations lead to a catalogue of TOAs which can
then be analyzed to search for GWs. 
A timing model describing all deterministic contributions to a
pulsar's TOAs (rotational frequency, spindown rate, dispersion measure
(DM), etc.) is iteratively fit with the analysis package,
\textsc{TEMPO2} \citep{tempo2-1,tempo2-2}. The difference between the
measured TOAs and the refined timing model prediction is the post-fit
timing residual, which constitutes the input data in our GW analysis. 

{\it GW analysis pipeline.--}
We use the signal modeling techniques described in
Ref.~\citep{epta-isotropy-2015} (from hereon L15). The
posterior probability of the model parameters, $\vec\Theta$, given the
concatenated post-fit timing residuals from all pulsars, $\vec{\delta t}$, is a
multivariate Gaussian:
\begin{equation}
p(\vec\Theta|\vec{\delta t})\!\propto \!p(\vec\Theta)\frac{\exp{\left(-\frac{1}{2}\vec{\delta
    t}^{\rm T}\mathbf{G}(\mathbf{G}^{\rm T}\mathbf{C}\mathbf{G})^{-1}\mathbf{G}^{\rm T}\vec{\delta
      t}\right)}}{\sqrt{{\rm det}{[2\pi
    (\mathbf{G}^{\rm T}\mathbf{C}\mathbf{G})]}}}        
    \end{equation}
where $p(\vec\Theta)$ is the prior probability distribution of model
parameters, and projecting all quantities with the matrix $\mathbf{G}$
marginalizes this posterior probability over all timing model
parameters (see Ref.\ \cite{van-haasteren-levin-2012}). The covariance of the
{\it pre-fit} timing residuals is defined as $\mathbf{C} = \mathbf{C}_{\rm red}+\mathbf{N}$ where
$\mathbf{C}_{\rm red}$ includes the SGWB, intrinsic pulsar red noise, and DM variation
components, while $\mathbf{N}$ denotes all white-noise
components. This red covariance is decomposed in terms of a low-rank
approximation such that $\mathbf{C}_{\rm red} =
\mathbf{F}\bm{\varphi}\mathbf{F}^{T}$, where $\mathbf{F}$ is a
block-diagonal matrix of Fourier basis vectors, and
$\bm{\varphi}$ is a spectral covariance matrix \citep{lentati-spectrum-2012,van-vallis-2014a,van-vallis-2014b}. Intrinsic red-noise and
 SGWB are expanded in the same Fourier basis, while the DM-variation
signal is expanded in basis-functions which differ only by an extra
multiplicative factor of $\propto 1/\nu_o^2$, where $\nu_o$ is the observing
frequency of the pulses. 
The matrix $\mathbf{N}$ is diagonal, with entries given by the squared
TOA errors which have been corrected by previous single pulsar
analyses \citep{epta-isotropy-2015}. We apply a multiplicative factor to all
error bars of a given pulsar (referred to as the GEFAC parameter)
which is searched over here.

The matrix $\mtrx{\varphi}$ has band-diagonal structure, since Fourier modes between different pulsars may be correlated due to the presence of a SGWB or correlated noise. Therefore
\begin{equation}
\![\mtrx{\varphi}]_{a i,b j}\!\!=\! \Gamma_{ab}\rho_i\delta_{ij} \!\!+\! \epsilon_i\delta_{ij} \!+\!\eta_i\delta_{ab}\delta_{ij}\!+\! \kappa_{a i}\delta_{ab}\delta_{ij}\!
\end{equation}
where $i$ and $j$ index the discretely sampled signal or noise frequencies in our
analysis of pulsar TOAs; $\rho=h_c(f)^2/(12\pi^2f^3T_\mathrm{max})$ is the power spectrum
of the SGWB, with $T_\mathrm{max}$ equal to the timing baseline of the
PTA; $\epsilon$
is the spectrum of a completely correlated red-noise process which may
result from modeling inaccuracies due to drifts in the
observatory and global time standards; $\eta$ is the spectrum of a common, but
uncorrelated, red-noise process which may originate from common
physical processes inside the neutron stars, see e.g.\ Refs.\ \citep{shannon-cordes-2010,hobbs2010}; and $\kappa_{a}$ is the individual
red-noise and DM-variation spectrum for pulsar $a$. All these spectra are modeled with
power laws,
$(A^2/12\pi^2T_\mathrm{max})(f_n/{\rm yr^{-1}})^{2\alpha-3}\;{\rm yr^2}$, where
$A=A_h$ for the SGWB; $\alpha$ is a spectral index which equals $-2/3$ for the SGWB;
and $f_n$ are the $n$ frequencies at which we sample the spectra of
red-noise processes, where in this analysis $n=50$.

The ORF, $\Gamma_{ab}$, is the average of the overlap of
the pulsars' antenna response functions, $F_a^A(\hat\Omega)$ (see
e.g.\ Refs.~\citep{TaylorGair:2013, MingarelliEtAl:2013}), over GW
propagation directions $\hat\Omega$, and weighted by the SGWB angular power
distribution, $P(\hat\Omega)$:
\begin{equation} 
\label{eq:GammaDef}
\Gamma_{ab} \!=\!
\frac{3}{8\pi}(1+\delta_{ab})\!\!\int_{S^2}\!\!\! d\hat\Omega\;P(\hat\Omega)\sum_{q}F_a^{q}(\hat\Omega)F^{q}_b(\hat\Omega)\, ,
\end{equation}
where $q$ labels the $\{+,\times\}$ GW polarization. An excess in $P(\hat\Omega)$ in a
particular region of the sky may indicate a particularly bright
single source, or a hotspot of several sources
\citep{cornish-sesana-2013,babak-sesana-2012,MingarelliEtAl:2013,TaylorGair:2013}. In
the following we decompose the SGWB angular distribution such that $P(\hat{\Omega}) \equiv \sum^{l_\mathrm{max}}_{l=0}\sum^{l}_{m=-l}c_{lm}Y_{lm}(\hat{\Omega})$,
with normalisation $\int_{S^2}P(\hat{\Omega})d\hat{\Omega} =
4\pi$, where $Y_{lm}$ are the real spherical harmonics. Inserting this decomposition into Eq.\
(\ref{eq:GammaDef}) and proceeding as in Ref.\
\citep{MingarelliEtAl:2013}, we expand $\Gamma_{ab}$ into a sum over
anisotropic ORFs, $\Gamma_{lm}^{(ab)}$, with associated weights, $c_{lm}$, to be constrained
by the analysis and which characterize the SGWB angular power distribution. 
We
note that the leading function in this expansion, $c_{00}\Gamma^{(ab)}_{00}$, corresponds to the ORF applicable to the monopole
moment of $P(\hat\Omega)$ (also known as the
Hellings and Downs curve \citep{hellings-downs-1983}). Hence, current
analysis strategies for isotropic SGWBs emerge from our fully general
anisotropy framework as a special case.

Each pulsar has $5$ stochastic parameters to be constrained in a
Bayesian analysis: intrinsic red noise $(A,\alpha$), DM-variation $(A,\alpha)$, and
a GEFAC parameter. The fully correlated red-noise component,
$\epsilon$, will contribute $2$ power-law parameters, as will the
common, uncorrelated process, $\eta$. The spectrum of the SGWB
is modeled with a fixed slope of $-2/3$ and an amplitude, $A_h$ to be constrained, while ${l_{\rm max}>0}$ analyses will
include ${[(l_{\rm max}+1)^2-1]}$ additional parameters. The
priors on parameters are: $\log_{10}A\in U[-20,-10]$,
$\alpha \in U[-2.0,1.5]$, ${\rm GEFAC}\in U[0.1,10.0]$. The $c_{lm}$
coefficients are constrained by a prior requiring the implied
distribution of GW power to be positive at all sky-locations
\citep{TaylorGair:2013}, called the \textit{physical
  prior}. The prior on $A_h$ is treated separately from
other red-noise components, and is uniform in the range
$[10^{-20},10^{-10}]$. Applying a uniform prior on $A_h$ with
logarithmic priors on the amplitudes of all other red components will provide
the most conservative upper-limits on the strain-spectrum amplitude of
the SGWB.

{\it Results.}-- We parametrize the angular distribution of the SGWB down to the angular resolution of the PTA. The most
anisotropic SGWB signal is one dominated by a single source. Hence the angular
resolution, and thus $l_{\rm max}$, is a function of the number of
pulsars, $N_\mathrm{psr}$, which significantly contribute to a single-source
detection, and the SNR of that detection~\cite{SesanaVecchio:2010}. \citet{SesanaVecchio:2010}
find that the angular resolution of a PTA for a resolvable GW
source is $\Delta \Omega \propto 50 \left(50/N_\mathrm{psr} \right)^{1/2} \left(10/\mathrm{SNR }\right)^2 \mathrm{deg}^2$, 
and this resolution sets an upper bound on $l$ via
$l=180/\theta$, where
$\theta=\sqrt{\Delta\Omega}$~\citep{GairEtAl:2014}. 
We analyse a subset of the six best pulsars in the EPTA \citep{paperI} which
encapsulate $\sim 95\%$ of the full-array SNR in simulated continuous
GW searches \cite{epta-cw-2015}: PSRs J$0613$$-$$0200$,  J$1012$$+$$5307$, J$1600$$-$$3053$, J$1713$$+$$0747$,
  J$1744$$-$$1134$, J$1909$$-$$3744$, where $T_{\rm max}=17.7$ years and the GW
frequencies with which we characterize red-noise components are
${\in\left[1/T_{\rm max}=1.79\;\mathrm{nHz}, 50/T_{\rm
      max}=89.7\;\mathrm{nHz}\right]}$. Hence, in our array subset $l_\mathrm{max}\lesssim 4$.
Carrying out
searches with the noise characteristics of each pulsar fixed, we find
the upper limits on the strain amplitude remain consistent whether we
analyze this six-pulsar subset or the full
array. 
Including more pulsars of comparably high timing quality would contribute a larger number of
pulsar-pairs [$N_{\rm pairs} = N_{\rm psr}(N_{\rm psr}-1)/2$], which would serve to increase the SNR and resolving
power ($l_{\rm max}$)
of any
 search for anisotropy. 
 This comes at the
cost of longer likelihood evaluation times, making the systematic study presented here currently intractable. Our goal is to
provide the first constraints on anisotropy in the SGWB via a
systematic study with current techniques -- we do so
with the $15$ distinct pulsar-pairings afforded by a six-pulsar
array. 
All analysis is performed with
parallel-tempering Markov chain Monte Carlo (MCMC) analysis.
\begin{figure*}
  
   \incgraph{0}{0.49}{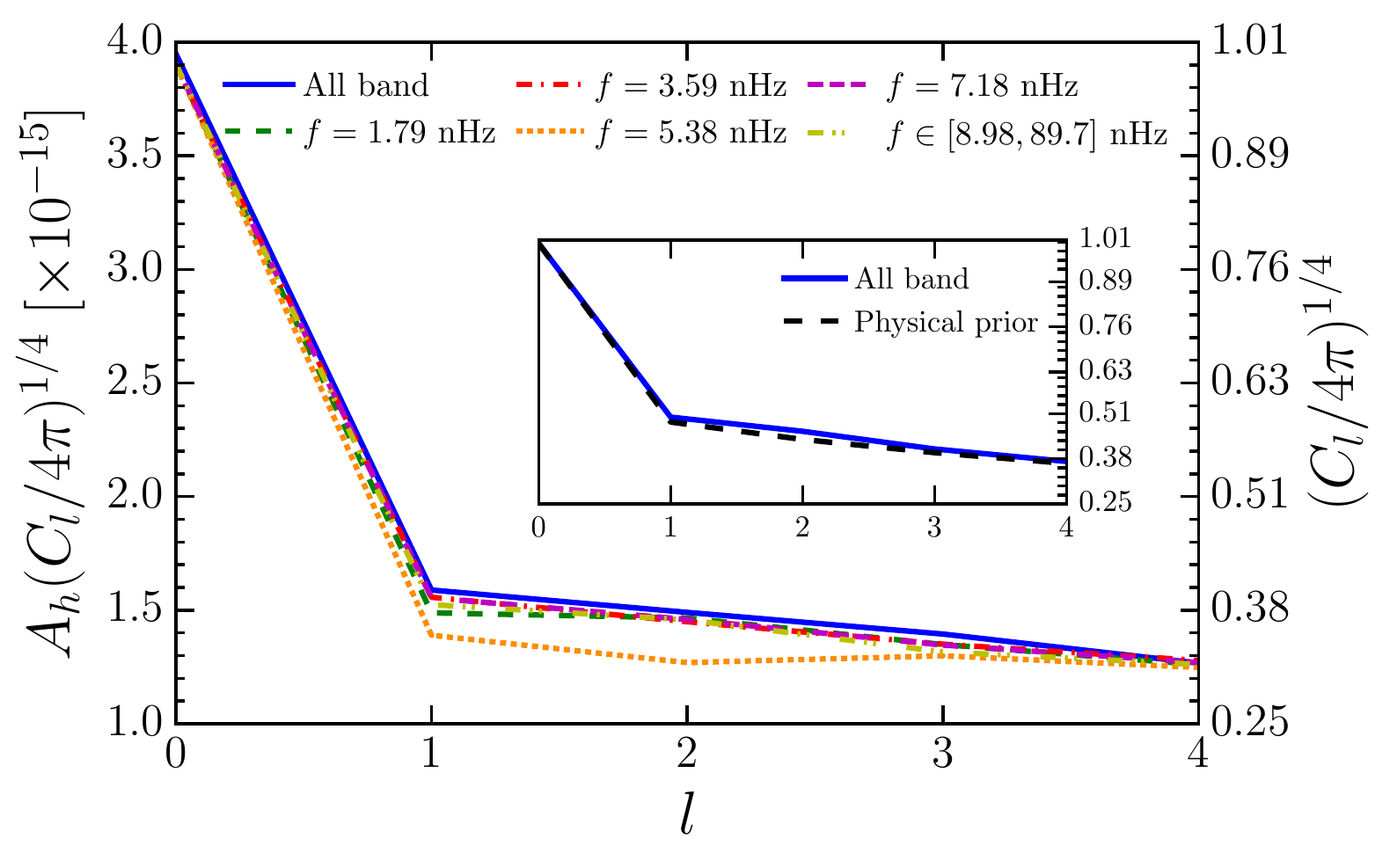}\hspace{1mm}
 \incgraph{0}{0.49}{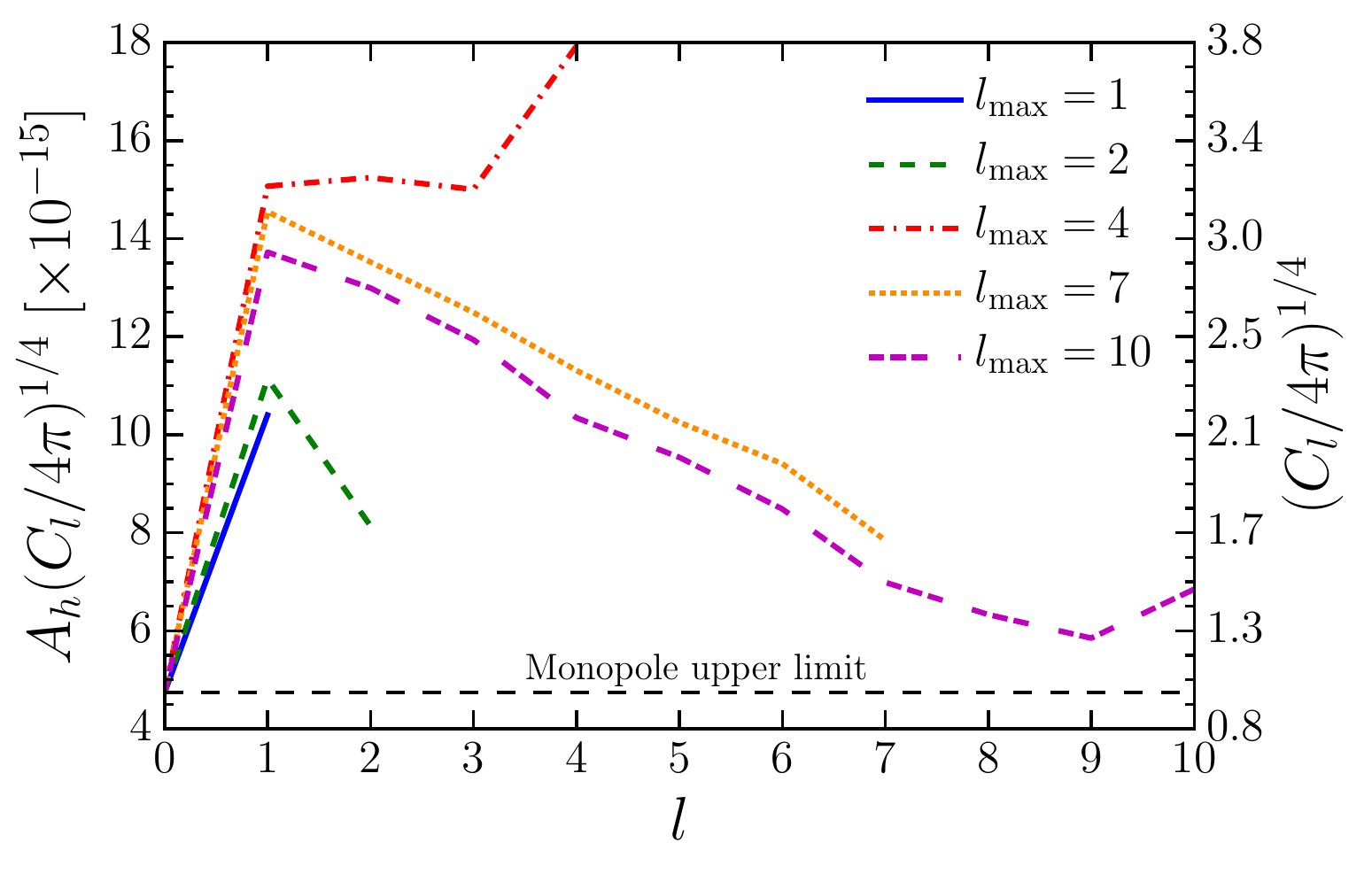}
   \caption{$95\%$ upper limits on the
     strain amplitude, where $C_l  = \sum_{m=-l}^l|c_{lm}|^2/(2l+1)$. \textit{Left}:
     \textit{all-band anisotropy parametrization} and
     \textit{frequency-dependent parametrization (ii)}. The right axis is the ratio of the
     upper limit to the monopole. The inset figure shows 
     $95\%$ upper limits on $(C_l/4\pi)^{1/4}$
     which are marginalized over the strain amplitude for the
     \textit{all-band anisotropy parametrization} and a constant
     likelihood analysis. Our limits reflect the constraints of the
     physical prior. \textit{Right}: \textit{all-band
       anisotropy parametrization}, where the $c_{lm}$ values are
     obtained by mapping %from the Bayesian MCMC-sampled
     cross-correlation values to the spherical harmonic basis, without
     physical prior rejection. } 
\label{fig:physprior_upper}
 \end{figure*}

 \begin{figure} [b]
\hspace{-0.2in}
  \centering
   \incgraph{0}{0.5}{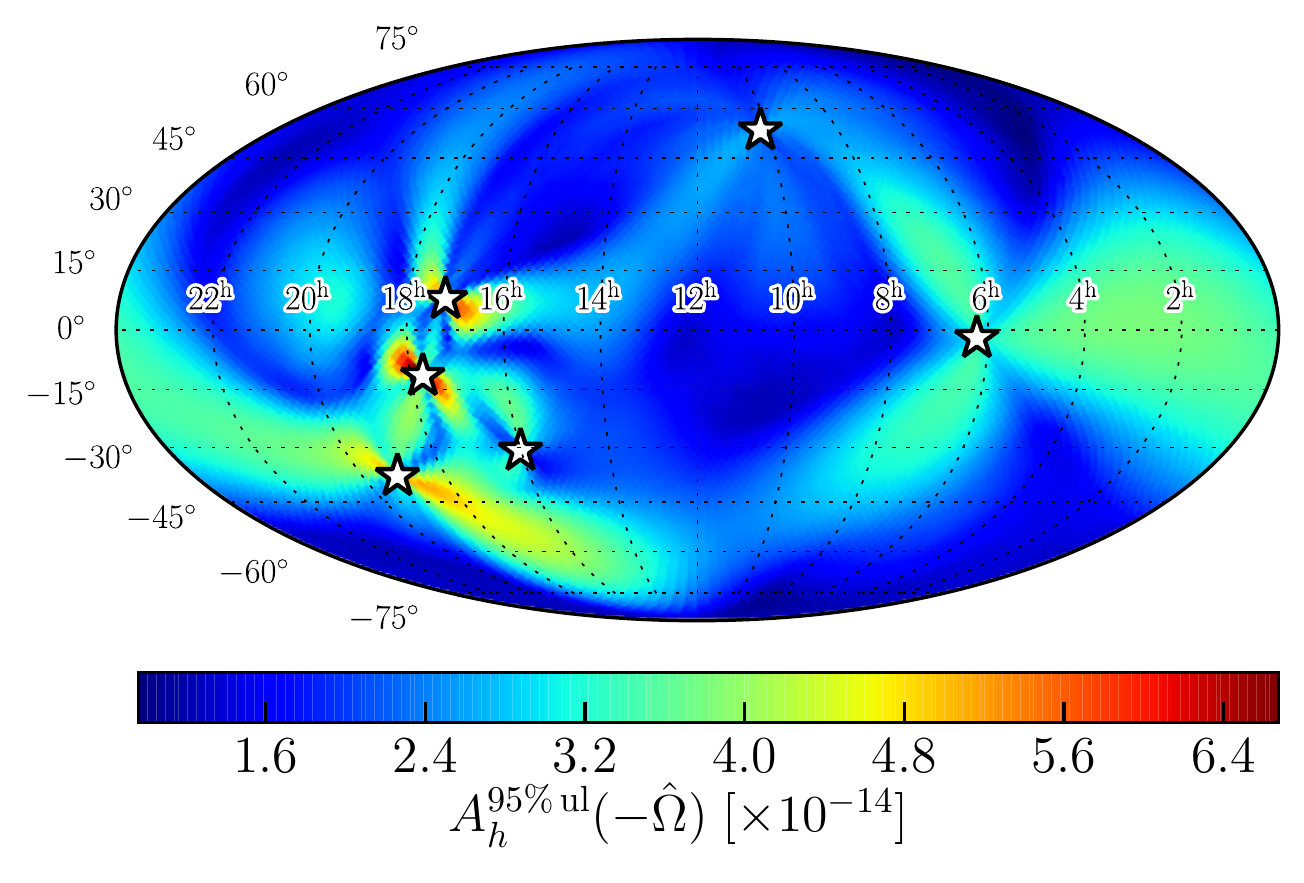}
   \caption{$95\%$ upper limits on the GW strain amplitude in each pixel. These limits are obtained by mapping from the Bayesian MCMC-sampled
     cross-correlation values to a pixelated ORF
   basis (${N_\mathrm{pix}=12288}$). White stars show the pulsar locations. } 
\label{fig:pixelbasis_upper}
 \end{figure}

The $95\%$ upper limits on $A_h$ from our
analyses are shown in Table
\ref{Table:LimitTable}. Firstly, we perform searches with a single set of anisotropy
coefficients $c_{lm}$ across the entire band, which we call the (\textit{All-band
  anisotropy parametrization}, cf. Table \ref{Table:LimitTable}). We also
perform frequency-dependent searches by parametrizing each frequency
with independent $c_{lm}(f)$ coefficients. We split our
band into $5$ equal sub-bands ($\Delta f = 17.9$ nHz), and independent $c_{lm}(f)$
coefficients constrained in each, called the \textit{Frequency-dependent anisotropy parametrization (i)}. 
Finally, motivated by the results of
L15 -- where most of the SGWB constraints were found to come from the lowest
three frequencies-- we apply independent $c_{lm}(f)$ coefficients to
the lowest four frequencies in our analysis ($f=[1,2,3,4]/T_\mathrm{max}=[1.79, 3.59, 5.38,
7.18]$ nHz), with the remainder of the
band ($f=[5,\ldots,50]/T_\mathrm{max}=[8.98,\ldots, 89.7]$ nHz) parametrized by a single set of coefficients. This is
reported as \textit{Frequency-dependent anisotropy parametrization (ii)}.
The recovered
upper limit does not deteriorate through the increased number of
parameters in our higher multipole searches. 
The monopole upper limits do not precisely match those found in
L15 due to several variations in the analysis specifics,
namely: $(i)$
our prior on the amplitude of red-noise components is uniform in
logarithm of the amplitudes which provides a
more conservative upper limit on the SGWB strain amplitude; $(ii)$
we do not consider Solar System ephemeris errors in our correlated
noise modeling; and $(iii)$ we employed a pure time-domain likelihood in the initial
single pulsar analysis to correct the TOA errors in each pulsar. Hence, our monopole upper limits are higher than in
L15 by $\sim
1\times 10^{-15}$. However, moving beyond the first analysis presented
here, our more general anisotropy framework can be easily incorporated into
all existing and planned pipelines to become a standard toolset, since it recovers the isotropic
SGWB constraints as a special case.
The upper limits on the strain-amplitude in each anisotropic multipole
of the search are shown in the left panel of Fig.\
\ref{fig:physprior_upper}, where the constraints are
entirely dominated by the restrictions imposed on the $c_{lm}$s
by the physical prior. Our data are not informative enough to update
the prior knowledge we have about the anisotropy of the GW sky.

Rather than impose a specific decomposition of the SGWB sky during
sampling, we can recover the cross-correlation values between
pulsar pairs and map these to a chosen basis in post-processing. We
perform a Bayesian search for the distinct elements of the Cholesky
factor of the residual cross-correlation matrix, which ensures
positive-definiteness of the final matrix
\cite{pinheiro1996unconstrained,lentati-spectrum-2012}.
After sampling we define a mapping
between the coefficients of the ORF in a particular basis, $\vec{c}$, and the
cross-correlation values, $\vec\Gamma$, such that $\vec\Gamma =
{\mathbf{H}}\vec{c}$. A single row of the matrix $\mathbf{H}$ will
have entries corresponding to the ORF between pulsars $a$ and $b$
evaluated for all basis terms. In the spherical-harmonic basis, such a
row would consist of $\left( \Gamma^{(ab)}_{00} \Gamma^{(ab)}_{1-1}
  \cdots \Gamma^{(ab)}_{lm} \right)$, and for a pixel basis this is
$\left( \Gamma^{(ab)}_{\hat\Omega_1} \Gamma^{(ab)}_{\hat\Omega_2}
  \cdots \Gamma^{(ab)}_{\hat\Omega_N} \right)$. Having recovered
posterior samples of the vector $\vec\Gamma$, we map these to
samples of $\vec{c}$ via $\vec{c}=\mathbf{H}^+\vec\Gamma$, where
$\mathbf{H}^+$ corresponds to the Moore-Penrose pseudo-inverse of the matrix
$\mathbf{H}$ \citep{dresden1920,PSP:2043984}. 
The results for mappings to the spherical-harmonic basis with varying
$l_{\rm max}$ are shown in Fig.\
\ref{fig:physprior_upper}(right). The data support such strong anisotropy signatures in
this model because the joint-posterior in the cross-correlation values
are consistent with essentially the entire range of $[-1,1]$, which when
mapped to a spherical-harmonic ORF-basis leads to large $c_{lm}$
values. There is nothing to penalize these large anisotropy
coefficients, which lead to highly anisotropic (and possibly negative) GW
power distributions and would otherwise be restricted by
the physical prior. This supports to our claim that the
constraints in Fig.\ \ref{fig:physprior_upper} (left) are prior-dominated.

We also map our recovered cross-correlation samples to a pixel basis
with $12288$ equal-area pixels on the sky. We supplement our mapping
with the additional normalization constraint that $\int_{S^2}P(\hat\Omega)d\hat\Omega\approx\sum_{i=1}^{N_{\rm pix}}
c(\hat\Omega_i)\Delta\hat\Omega_i=4\pi$. The resulting SGWB power in each pixel
is marginalized over all other pixels and truncated to obtain the
positive $1$D-marginalised power PDF before it is integrated over to
obtain the upper limit on the strain-amplitude in that pixel. The
result is shown in Fig.\,\ref{fig:pixelbasis_upper}, where we see the distinctive overlapping antenna
patterns of the pulsars mapping out the sensitivity of the
PTA to the background strain-amplitude. The constraints on $A_h$ from
each pixel are quite poor, and in some cases are more than an order of
magnitude worse than the all-sky upper limit. As we decrease the
resolution of the pixelation the constraints in each pixel become
tighter, until we reach the limit of one pixel, which recovers the
usual all-sky upper-limit. Figure \ref{fig:pixelbasis_upper} can also
help to explain the results in the right panel of Fig.\
\ref{fig:physprior_upper}, where we see that the distribution of
pulsars in our array leads to the sub-optimal overlapping of the
antenna response functions,
which in turn causes insensitivities around the $4$ clustered pulsars and on
large angular scales. Hence, we will lack sensitivity to large angular
scale anisotropy ($l\sim 1$), which is reflected in the right panel of
Fig.\,\ref{fig:physprior_upper}. Moreover, this sensitivity map illustrates the
importance of timing pulsars from all over the sky to ensure a more
uniform sensitivity to GW strain, which will be possible through
international collaborations such as the International PTA\citep{iptareview2013}.

{\it Conclusions.}-- Our analyses suggest that this dataset is not
informative enough to update our prior knowledge of the angular distribution
of the nanohertz SGWB. Using a prior which enforces a positive SGWB
distribution, we find that the $95\%$
upper limit on the strain amplitude in multipoles of the background
distribution with $l>0$ is $\lesssim 40\%$ of the monopole strain. No
evolution of these upper-limits as a function of GW frequency is found
since the constraints are a reflection of the prior. Additionally, we
can recover the joint posterior distribution of the cross-correlation
values between pulsar pairings, and subsequently map these to a
spherical-harmonic or pixel ORF-basis. With the only constraint being
positive-definiteness of the cross-correlation matrix, the
strain-amplitude in $l>0$ multipoles is $\lesssim 400\%$ of the
monopole value. The
strain-amplitude upper-limits as a function of location on the sky reflect the
overlapping antenna pattern behavior of the full PTA, where the limits can often
be more than an order of magnitude worse than the all-sky limit. A
full description of all techniques employed here, and their efficacy, will
be provided in a follow-up methods paper.

Forthcoming advanced radio instruments such as the
Five-hundred-metre Aperture Spherical Radio Telescope \citep[FAST,][]{fast2011,fast2014},
MeerKAT \citep{meerkat2009}, and the Square
Kilometre Array \citep[SKA,][]{janssen2015} will enhance the detection and
inference prospects for anisotropic GW skies by detecting
large numbers of millisecond pulsars and timing them to unprecedented
precision. Upcoming studies will investigate how we can combine galaxy catalogues with frequency-dependent maps of the
nanohertz GW sky to probe whether the strain budget is being dominated by a few bright nearby sources, or is
more diffuse. We hope that the work presented here, together with these future
studies, will provide important insights into the demographics,
evolution, and assembly of SMBHBs not accessible by any other means.

\begin{table}
\centering
\begin{tabular}{lc|c|c}
\hline
$l_{\rm max}$        & $A_h$; all-band $c_{lm}$       & $A_h$;
$c_{lm}=c_{lm}(f)^{(i)}$        &  $A_h$; $c_{lm}=c_{lm}(f)^{(ii)}$  \\
\hline
\hline
$0$ & $3.94\times 10^{-15}$ & N/A                              & N/A \\
$1$ & $4.09\times 10^{-15}$ & $4.06\times 10^{-15}$ & $4.06\times 10^{-15}$ \\
$2$ & $4.06\times 10^{-15}$ & $4.07\times 10^{-15}$ & $4.02\times 10^{-15}$ \\
$3$ & $4.06\times 10^{-15}$ & $3.98\times 10^{-15}$ & $4.01\times 10^{-15}$ \\
$4$ & $4.03\times 10^{-15}$ & $3.95 \times 10^{-15}$ & $3.99\times 10^{-15}$\\
\hline
\end{tabular}
\caption{$95\%$ upper limits on SGWB strain amplitude $A_h$. The first
  column is the all-band anisotropy parametrization, the second and
  third correspond to the frequency-dependent anisotropy parameterizations
  $(i)$ and $(ii)$ respectively, described in the text.}
\label{Table:LimitTable}
\end{table}

{\it Acknowledgments.}-- This work was carried out under the aegis of the EPTA. Part of this work is based on observations with the 100-m telescope of the Max-Planck-Institut f{\"u}r Radioastronomie (MPIfR) at Effelsberg. The Nan{\c c}ay radio Observatory is operated by the Paris Observatory, associated to the French Centre National de la Recherche Scientifique (CNRS). We acknowledge financial support from `Programme National de Cosmologie and Galaxies' (PNCG) of CNRS/INSU, France.   Pulsar research at the Jodrell Bank Centre for Astrophysics and the observations using the Lovell Telescope is supported by a consolidated grant from the STFC in the UK. The Westerbork Synthesis Radio Telescope is operated by the Netherlands Institute for Radio Astronomy (ASTRON) with support from The Netherlands Foundation for Scientific Research NWO. 
SRT acknowledges the support of the STFC and the RAS. This research
was in part supported by SRT's appointment to the NASA Postdoctoral
Program at the Jet Propulsion Laboratory, administered by Oak Ridge
Associated Universities through a contract with NASA. CMFM
was supported by a Marie Curie International Outgoing Fellowship
within the 7th European Community Framework Programme. This research
was performed in part using the Zwicky computer cluster at Caltech
supported by NSF under MRI-R2 award no.\ PHY-0960291 and by the Sherman Fairchild Foundation. This work was
in part performed using the Darwin Supercomputer of the University of
Cambridge High Performance Computing Service
(http://www.hpc.cam.ac.uk/), provided by Dell Inc. using Strategic
Research Infrastructure Funding from the Higher Education Funding
Council for England and funding from the STFC. 
AS and JG are supported by the Royal Society. LL was supported by a
Junior Research Fellowship at Trinity Hall College, Cambridge
University. SAS acknowledges funding from an NWO Vidi fellowship (PI
JWTH). RNC acknowledges the support of the International Max Planck
Research School Bonn/Cologne and the Bonn-Cologne Graduate School.
KJL is supported by the National Natural Science Foundation of China
(Grant No.11373011).  RvH is supported by NASA Einstein Fellowship
grant PF3-140116.  JWTH acknowledges funding from an NWO Vidi
fellowship and ERC Starting Grant `DRAGNET' (337062).   PL
acknowledges the support of the International Max Planck Research
School Bonn/Cologne. KL acknowledges the financial support by the
European Research Council for the ERC Synergy Grant BlackHoleCam under
contract no. 610058. SO is supported by the Alexander von Humboldt
Foundation. The authors also acknowledge support of NSF Award PHY-1066293 and the 
hospitality of the Aspen Center for Physics. Copyright \copyright\; 2015. All rights reserved.

\bibliography{references}

\end{document}